# Colloidal Plasmonic Titanium Nitride Nanoparticles: Properties and Applications

*Urcan Guler[†,‡], Sergey Suslov[§], Alexander V. Kildishev[†,‡], Alexandra Boltasseva[†,‡,||], Vladimir M. Shalaev[\*,†,‡]*

[†] School of Electrical & Computer Engineering and Birck Nanotechnology Center, Purdue University, West Lafayette, IN 47907, USA

[‡] Nano-Meta Technologies Inc., 1281 Win Hentschel Boulevard, West Lafayette, IN 47906, USA

[§] Birck Nanotechnology Center, Purdue University, West Lafayette, IN 47907, USA

[||] DTU Fotonik, Department of Photonics Engineering, Technical University of Denmark, Lyngby, DK-2800, Denmark

ABSTRACT. Optical properties of colloidal plasmonic titanium nitride nanoparticles are examined with an eye on their photothermal via transmission electron microscopy and optical transmittance measurements. Single crystal titanium nitride cubic nanoparticles with an average size of 50 nm exhibit plasmon resonance in the biological transparency window. With dimensions optimized for efficient cellular uptake, the nanoparticles demonstrate a high photothermal conversion efficiency. A self-passivating native oxide at the surface of the nanoparticles provides an additional degree of freedom for surface functionalization.

Materials research has been the backbone in the field of plasmonics for the last few years. Unique properties supplied by alternative plasmonic materials have created an excitement to develop technologies for a broad range of applications. The wide range of carrier concentrations available from several material classes span the electromagnetic spectrum from the ultra-violet through the mid-infrared.[1] Additionally, the broad variety of materials studied in the field provide extra degrees of freedom to solve technological challenges thanks to the unique properties such as tunability, process compatibility, chemical stability etc.[2]

Transition metal nitrides, especially titanium nitride (TiN), have been studied for the last three decades due to the unique combination of their material properties.[3] Metallic behavior of TiN combined with its hardness and chemical stability have attracted attention in microelectronics research.[4] Adjustable optical properties and stoichiometry of TiN thin films were examined in the 1990s.[5] Optical characterization of TiN thin films was extensively performed later with the increasing interest in the field of plasmonics.[6-9] Improved properties with epitaxial TiN thin films have recently been demonstrated with a hyperbolic metamaterial and a plasmonic waveguide.[10,11] As a refractory plasmonic material, TiN holds the potential to solve critical issues associated with the softness and low melting points of plasmonic metals such as gold and silver.[12] In contrast to thin films, TiN nanoparticles have not been extensively studied until recently. Localized surface plasmons (LSPs) in TiN nanoparticles were first theoretically analyzed by Quinten,[13] while their first experimental demonstration was performed by Reinholdt et al.[14] In their work, cubic nanoparticles smaller than 10 nm with a broad size dispersion were fabricated through laser ablation, and plasmon resonance peaks centered around 730 nm wavelength were reported. Reinholdt et al. proposed the use of TiN nanoparticles as inorganic, stable color pigments. We have previously shown that TiN nanoparticles provide field enhancement comparable to that achievable with Au, with a broader peak spectrally positioned in the biological transparency window.[15] In a more recent study, we experimentally analyzed the absorption efficiencies of lithographically fabricated TiN and Au nanoparticles and showed that TiN nanoparticles can provide enhanced local heating in the biological transparency window when compared to identical Au samples.[16]

As a refractory plasmonic material, TiN delivers chemical stability at elevated temperatures. The high melting point and hardness of the material enables the use of plasmonic effects for applications where tough operational conditions are required.[12] Heat assisted magnetic recording,[17] solar/thermophotovoltaics,[18] and plasmon assisted chemical vapor deposition [19] are some of the applications that could take advantage of durable plasmonic TiN nanostructures. TiN is a chemically inert material widely used in CMOS and biomedical devices.[4, 20-23] Combined with the localized surface plasmon resonance (LSPR) peak located in the biological transparency window,[15] bio-compatibility of the material makes it a promising alternative for plasmonic photothermal therapy.[16] Consequently, the plasmonic properties of TiN colloidal nanoparticles are clearly important for future studies on the feasibility of the material for therapeutic applications. In this Letter, we examine the optical properties of aqueous TiN nanoparticle solution and show that the dipolar resonance matches well with the biological transparency window, suppressing the need for large nanoparticles and relatively complex systems of conventional metals that have been used so far. Transmission electron microscope (TEM) studies show that TiN nanoparticles are single crystalline with lattice constants in agreement with the literature values. The observed optical properties of TiN nanoparticles are of the same quality as those of the epitaxial thin films grown on c-sapphire substrates reported recently.[16] A thin native oxide layer, which can be removed or thickened at will,[24, 25] is observed at the surface of TiN nanoparticles. Surface oxide is a self-passivating layer with a thickness of 1-2 nm. The presence of a thin oxide layer could provide an additional degree of freedom for surface chemistry studies. Extensively studied oxide chemistry could be useful for surface functionalization where the inertness of TiN could be a problem. These results reveal the high potential of TiN as a plasmonic material for photothermal therapy. Further work is required in order to test the bio-compatibility and stability of nanoparticles as well as clearance from the body after treatment. The surface chemistry of TiN colloidal nanoparticles, with or without the native oxide layer, is an interesting research area.

Enhanced absorption properties of plasmonic nanoparticles are desired for applications such as plasmonic photothermal therapy where local heating is required. Nanoparticles can be delivered into a tumor region selectively, and a confined volume around the plasmonic nanoparticle can be efficiently heated via laser illumination at resonance wavelengths.[26] Accumulation of the nanoparticles and their selective heating results in the ablation of the tumor while damage to healthy tissue is kept minimal.[26, 27] One of the fundamental limitations of the method is the small penetration depth of light through biological tissue, usually on the order of a few centimeters. The near infrared region of the electromagnetic spectrum is known as the biological transparency window due to the reduced absorption of light in the region between wavelengths 700 nm and 1000 nm, enabling deeper penetration and more efficient power delivery to the region of interest. Thus, it is of high importance to design nanoparticles with plasmonic resonances in the biological transparency window. Gold has been the primary material for use in plasmonic photothermal therapy due to several factors. Bio-compatibility, high plasmonic performance and well-studied surface chemistry are among the top reasons for the frequent use of Au. However, the dipolar LSPR peak obtained from an Au nanoparticle is located around a wavelength of 550 nm, which is away from the biological transparency window. This spectral mismatch problem can be solved by using nanoparticles with relatively larger sizes, or engineered geometries. Hirsch et al. demonstrated a smart solution by using Au nanoshells in order to achieve plasmonic resonance in the biological transparency window at the expense of relatively large particle sizes exceeding 100 nm.[28, 29] However, the large size leads to additional complications such as limited cellular uptake and more difficult clearance from the body. It has been shown that cellular uptake is more efficient for nanoparticle sizes around 50 nm.[30] Biodegradable structures were recently suggested as a solution to the clearance problem.[31] In their work, Huang et al. demonstrated the use of closely packed 26 nm Au nanoparticles, with a total cluster size of 200 nm, as plasmonic antennas for photothermal therapy and reported improved clearance (most likely due to the dissociation of the nanoparticle assemblies after laser irradiation). However, this concrete step towards the improved clearance brings new questions such as the stability of the nanoparticle assembly and controllability of the biodegradation process. In another recent study, Goodman et al. reported the instability of hollow Au-Ag nanoshells, which provide shell structures with dimensions smaller than 100 nm, during *in-vivo* applications.[32] Another approach to red-shift the plasmonic resonance is to engineer the nanoparticle aspect ratio. Gold nanorods can be used as simpler geometry particles for theranostics.[33] However, surfactants originating from the synthesis of the particle and suppressed cellular uptake due to the increased aspect ratio are the problems that need to be addressed. [30, 34, 35] Carbon nanotubes exhibit a broad absorption band in the near infrared region making them interesting for photothermal therapy; however, compared to plasmonic nanoparticles, absorption efficiencies are relatively low. [26]

Titanium nitride exhibits optical properties similar to those of Au and has dielectric permittivity zero crossover in the visible wavelength range, making it plasmonic in this spectral region. Previously we have calculated that the scattering efficiencies of spherical TiN nanoparticles are comparable to those achieved in Au nanoparticles, and the dipolar peak of TiN is positioned in the near infrared region making it promising for theranostic applications.[15] Recently, experimental verification of the higher absorption efficiencies obtained from lithographically fabricated TiN nanoparticles compared to identical Au structures were reported.[16] Combined with the bio-compatibility of TiN, high absorption efficiency of simple-geometry nanoparticles is of high interest.[22] Titanium nitride nanoparticles in powder form can be fabricated by using nitridation of Ti or $TiO_2$ at high temperatures, laser ablation of Ti or TiN targets, mechanical alloying, microwave plasma technique, vapor synthesis, reduction-nitridation, and other techniques.[14, 24, 36-42] In this work, aqueous solutions are prepared with commercially available TiN powders (PlasmaChem) with an average nanoparticle size of 50 nm. Optical transmittance measurements from aqueous solutions are performed with a Lambda 950 Vis/NIR Spectrophotometer (Perkin Elmer). Nanoparticle size distribution is determined by dynamic light scattering (DLS) technique with a Zetasizer (Malvern Instruments). FEI Titan 80/300 Environmental TEM equipped with a Tridiem 863 Gatan Imaging Fiter (GIF) is used for electron microscopy. Samples for TEM characterization are prepared on Lacey Carbon Films (Ted Pella) to avoid substrate signal.

Energy filtered TEM (EF-TEM) images of TiN nanoparticles dispersed on a Lacey Carbon membrane are presented in Figure 1. The electron beam passing through the nanoparticle sample experiences energy losses unique to the elements in the sample. Subsequent to traversing the sample, the electron beam is dispersed through a magnetic prism and specific energy windows are selected with an energy slit. Filtered electrons, with the energy loss characteristic to a particular element, can be used to form an image of the element of interest.[43] We investigate the distribution of nitrogen (N), titanium (Ti) and oxygen (O) in TiN nanoparticles with a slit width of 5 eV located around the energies 401, 456 and 532 eV, respectively. Figure 1 (a-b) shows the unit cell of a rock salt TiN crystal and a TEM image of the sample under test, respectively. The most stable form of TiN, cubic shape,[14, 44] is evident from the image. Figure 1 (c) shows the combined image of the elemental maps, while Figure 1 (d-f) shows the maps of each element under investigation. The presence of Ti and N is clear from the images. A native oxide is also observed on the surface of the nanoparticles. In accordance with the literature, the thickness of the oxide at room temperature is around 1-2 nm. The presence of the native oxide layer is of particular importance. First of all, if the oxide layer is not desired, it can be removed by nitridation at elevated temperatures.[24] On the other hand, we find it particularly important to have an oxide layer since the chemical inertness of TiN could be a potential problem for surface chemistry studies. A thin oxide layer at the surface of the nanoparticles could be used as an additional degree of freedom for surface modifications. In fact, wide use of titanium oxides as a pigment in food and personal care products resulted in intensive research on the chemistry of nanometer scale particles.[45] Well-established chemistry of titanium oxide nanoparticles would be useful for surface functionalization and particle dispersion of oxide coated TiN nanoparticles. Additionally, the thickness of the oxide layer could be controllably increased at elevated temperatures, if desired.[46]

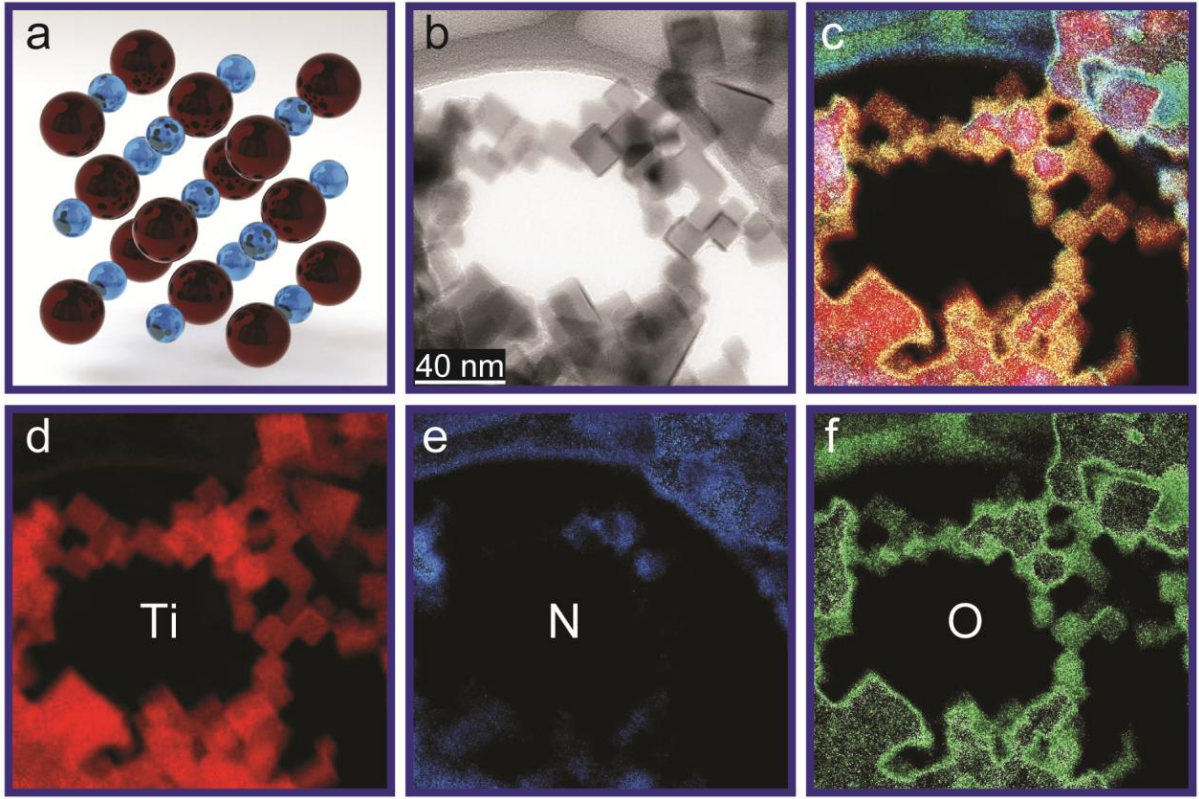

Figure 1: (a) Illustration of a rock salt TiN crystal unit cell. (b) TEM image of TiN nanoparticles with dominant cubic shapes following the rock salt lattice. (c) Combined elemental map of the same sample region. (d-f) Individual elemental maps of Ti, N and O, respectively. The self passivating native oxide layer at the surface of the nanoparticles is distinct with a 1-2 nm thickness.

The optical properties of TiN vary significantly depending on the stoichiometry and crystallinity of the structure. Single crystalline nanoparticles with minimal defects are desired for better oscillator quality and plasmonic performance. Figure 2 shows high-resolution TEM (HRTEM) images of a TiN nanoparticle, which reveals the cubic shape of the nanoparticle as well as the lattice planes. The thin oxide layer at the surface of the nanoparticle can be noticed from the HRTEM image as well. Figure 2 (b) is the illustration of the rocksalt structure with [100] lattice vector pointing out of the page plane. A fast Fourier transform (FFT) diffractogram obtained from the region shown in the inset is given in Figure 2 (c). Figure 2 (d) shows the simulated FFT diffractogram with the constants taken from the tabulated data.[47] The lattice constant of an individual nanoparticle is measured to be 4.22 Å, which is very close to the tabulated bulk value of 4.24 Å. Based on the experimental results showing the high crystalline quality of nanoparticles, we use a dielectric permittivity measured from lattice-matched, single crystalline TiN thin films in our simulations in this work.[16]

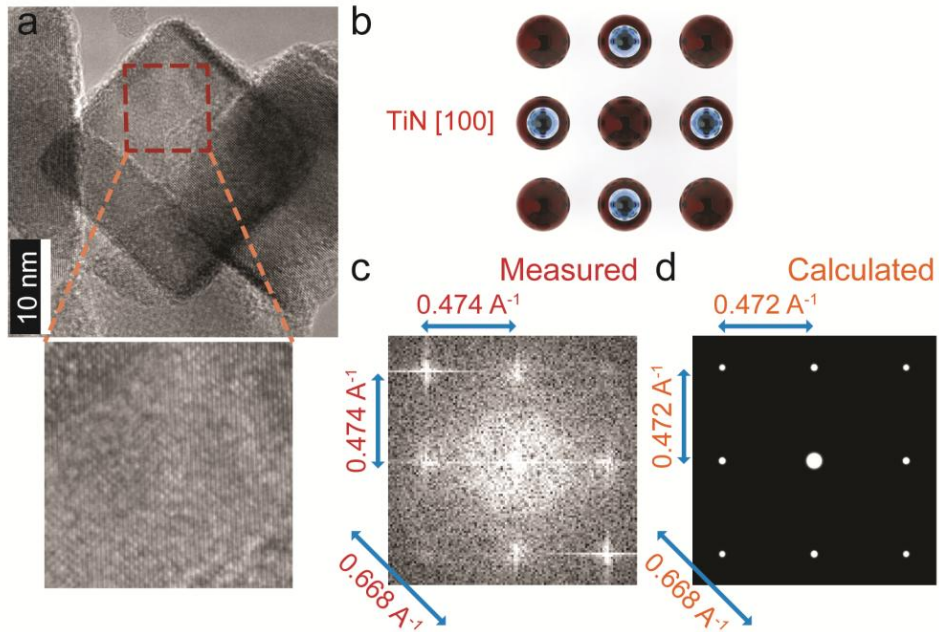

Figure 2: (a) HRTEM image of a single cubic TiN nanoparticle. The inset shows magnified image of the region designated with dashed rectangle. (b) The unit cell of rock salt TiN crystal with [100] lattice vector pointing out of the page plane. (c) FFT diffractogram of the designated region taken from the HRTEM image. (d) Calculated FFT pattern from tabulated data.[47] Measured lattice constant, 4.22 Å, is in good agreement with tabulated bulk value, 4.24 Å.

It has been demonstrated that lithographically fabricated TiN nanoparticles are efficient plasmonic absorbers in the biological transparency window.[16] However, lithographic nanostructures deposited under optimized growth conditions have limited use in biological applications. Colloidal TiN nanoparticles with similar plasmonic quality are essential for applications such as photothermal treatment. Among several materials with high plasmonic performance, we choose Au as the reference material due to its chemical stability and bio-compatibility. Due to their well demonstrated success in biological applications, we use nanoshell structures in order to compare the performance of colloidal TiN nanoparticles. Figure 3 shows the absorption cross-sections and efficiencies of TiN nanoparticles with sizes of 20, 50 and 80 nm. Gold nanoshells designed for resonance in the bio-transparency window are included as a reference. The inner and outer radii of the nanoshells were chosen to be 55 and 65 nm as these parameters are widely used in photothermal treatment.[28] Figure 3 (a) shows that absorption cross-sections comparable to Au nanoshells can be obtained with TiN nanoparticles of much smaller dimensions. For example, 80 nm TiN particles provide the same absorption cross-section peak with a volume 2.25 times less and cross-sectional area 2.07 times smaller than the nanoshell. 50 nm particles, providing approximately one third of the absorption cross-section, are 9.2 times smaller in volume and 5.3 times smaller in the cross-sectional area. As a result, the absorption efficiencies of TiN nanoparticles given in Figure 3 (b) exceed the efficiency of the reference nanoshell. 50 nm particles, which are found to have the optimum dimensions for efficient cellular uptake,[30] provide 50% better absorption efficiency when the peak positions are comparable and have slightly better efficiency at 800 nm wavelength. The broader resonance peak obtained from TiN nanoparticles is another advantage that relaxes the restrictions on particle size dispersion. It should be noted that smaller nanoparticle dimensions are desirable due to many critical requirements including efficient cellular uptake and clearance of particles after treatment.[30-32]

Scattering efficiency of TiN nanoparticles, on the other hand, is not superior to Au nanoshell. Here, the main reason is the significant size difference which favors scattering from a larger nanostructure.[48] It has also been shown that TiN nanoparticles are relatively deficient scatterers compared to identical structures made of Au.[15] Figure 3 (c) shows the scattering cross-section of TiN nanoparticles and Au nanoshells where a significant difference can be observed. The difference in the scattering efficiencies given in Figure 3 (d) is reduced, but the TiN peak values are still not comparable to those of Au nanoshells. Consequently, the use of TiN nanoparticles for local heating applications such as photothermal therapy and drug delivery seem very possible, applications based on the near-field enhancement would require careful optimization.

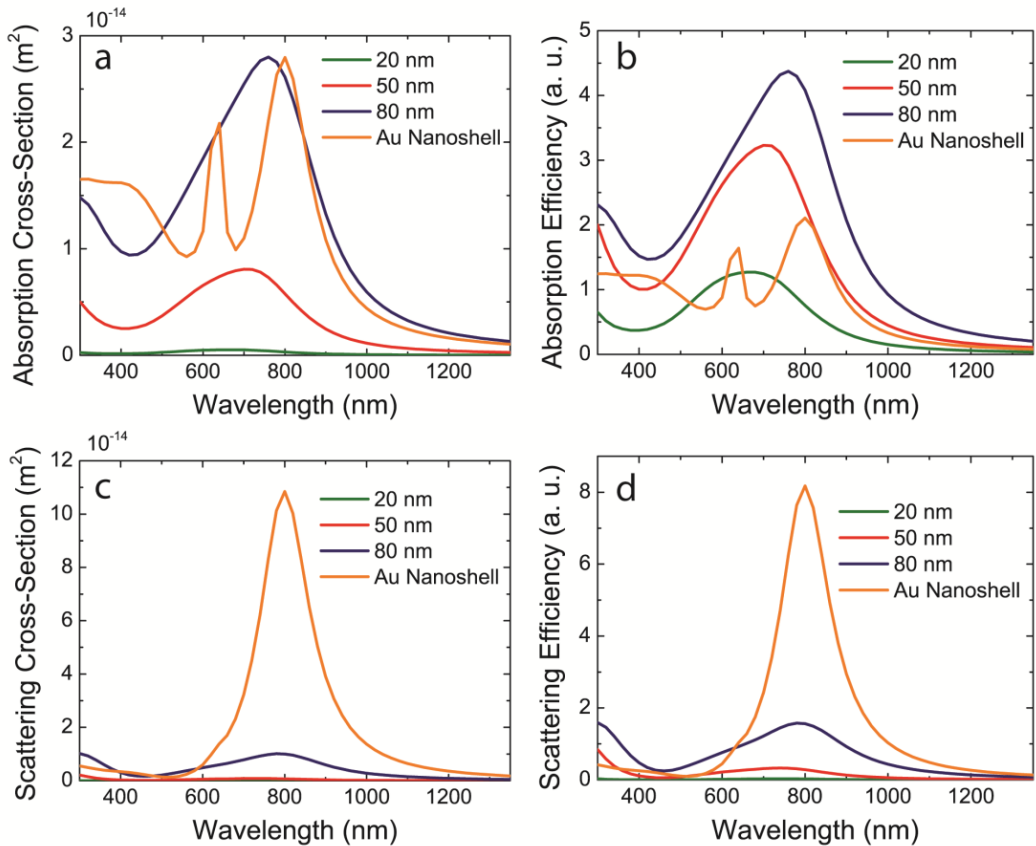

Figure 3: (a) Calculated absorption cross-section of TiN nanoparticles with sizes of 20, 50 and 80 nm compared to Au nanoshell with inner and outer diameters of 110 and 130 nm, respectively. (b) Absorption efficiencies of the same nanoparticles. Higher efficiencies in the biological transparency window obtained with smaller TiN nanoparticles would resolve problems arising from larger sizes of Au nanoshells. (c-d) Scattering cross-section and efficiencies of the same nanoparticles. Au nanoshell provides higher scattering efficiencies due to a larger particle size and more metallic behavior of the material.

We use optical transmittance data as an evidence of the plasmonic properties of colloidal TiN nanoparticles. The assumption that the dielectric permittivity of TiN nanoparticles are similar to that of the epitaxially grown thin films can also be checked by comparing experimental data and simulations. Figure 4 (a) shows the size dispersion of colloidal TiN nanoparticles dispersed in water. Dynamic light scattering reveals that the majority of the nanoparticles are smaller than 100 nm with a peak around 50 nm and a tail extends up to a few hundred nanometers that is likely due to clustering and marginal number of large nanoparticles. With a peak around 50 nm and shape aspect ratio of 1, the sample is suitable for efficient cellular uptake.[30] Figure 4 (b) shows the transmittance data of the sample along with the simulations of the aqueous solution consisting of mono-sized 50 nm particles and a solution with wider size distribution. Experimental data with a transmittance dip around 700 nm agrees well the simulation results presented in previous and the current work. Calculated transmittance data for a mono-sized sample exhibits a narrower peak with very accurate amplitude, verifying the accuracy of the assumption that nanoparticles are single crystalline with high purity. The deviation in the width of the peak is likely due to the broad size dispersion of the colloidal sample and clustering effects. A simple estimation with the addition of large nanoparticles provides results that resemble the experimental data. It is anticipated that future development of TiN synthesis and surface chemistry will uncover better control of size dispersion.

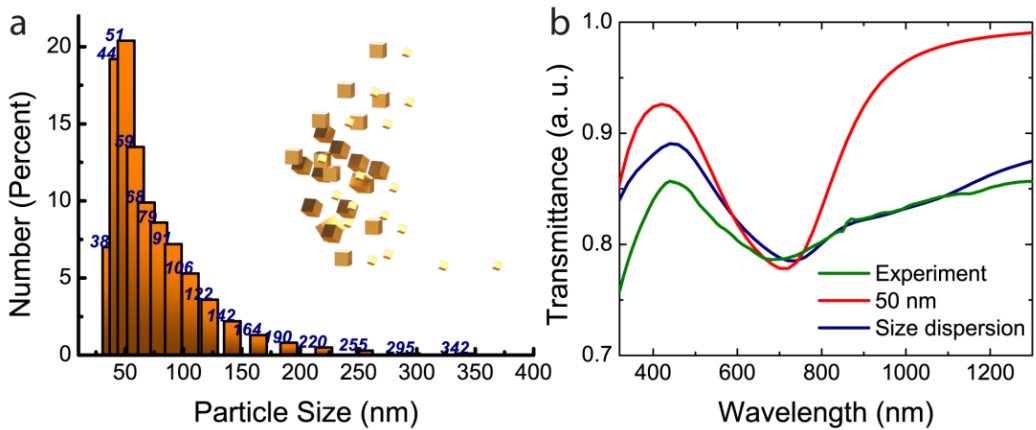

Figure 4: Dynamic light scattering measurement of the size dispersion for colloidal plasmonic TiN nanoparticles. Peak value around 50 nm is a desired size for more efficient cellular uptake. The long tail with marginal large dimensions is likely due to clustering effects. (b) Transmittance data taken from aqueous TiN sample (green) and calculated data for mono-sized 50 nm nanoparticle solution (red) and for a sample with marginal number of large nanoparticles (blue). Calculated transmittance with mono-sized nanoparticle solution agrees well on the peak spectral position and amplitude with the experimental data. Broad peak width can be reproduced by adding larger nanoparticles (up to 300 nm) to calculation.

Here, we present results of an initial study on plasmonic TiN colloidal nanoparticles with a focus on photothermal therapy and local heating applications. The results show that plasmonic TiN colloidal samples hold enourmous potential, hence requiring further studies involving the bio-compatibility at nanoscale, surface chemistry, and size dispersion control.

In conclusion, colloidal TiN nanoparticles exhibit plasmonic properties in the visible region and mimic the optical properties of recently developed crystalline TiN thin films. 50 nm particles provide absorption efficiencies better than larger Au nanoshell structures, leading to the opportunity of using smaller and simpler plasmonic nanoparticles for photothermal therapies. A thin native oxide shell formed on TiN structures is likely to provide an additional degree of freedom for surface functionalization of the nanoparticles.

ASSOCIATED CONTENT

AUTHOR INFORMATION

Corresponding Authors:

*Email: shalaev@purdue.edu.

Notes: The authors declare no competing financial interest.

ACKNOWLEDGMENT

We would like to thank Dr. Gururaj Naik for fruitful discussions and Nathaniel Kinsey for help in preparation of this manuscript. The authors acknowledge support from the following grants: ONR-MURI grant N00014-10-1-0942, ARO grant 57981-PH (W911NF-11-1-0359) and NSF MRSEC grant DMR-1120923.

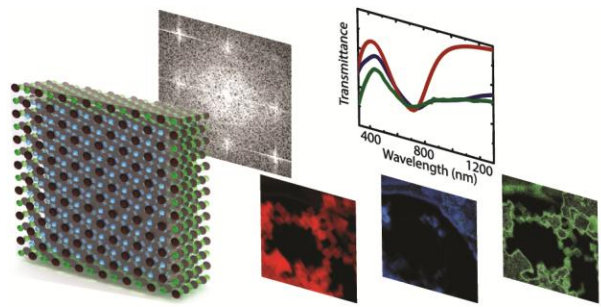

17